\documentclass[letterpaper,twocolumn,prl,aps,superscriptaddress,amsmath,amssymb,floatfix]{revtex4-1}
\usepackage{mathptmx}
\usepackage[utf8]{inputenc}
\usepackage{color}
\usepackage{amsmath}
\usepackage{amssymb}
\usepackage{graphicx}
\usepackage{esint}
\usepackage{ulem}
\usepackage[unicode=true,
 bookmarks=true,bookmarksnumbered=false,bookmarksopen=false,
 breaklinks=false,pdfborder={0 0 1},backref=false,colorlinks=true]
 {hyperref}
\hypersetup{
 linkcolor=magenta,urlcolor=blue,citecolor=blue,pdfstartview={FitH},hyperfootnotes=false}

\makeatletter

\DeclareFontEncoding{LGR}{}{}

\ProvideTextCommand{\~}{LGR}[1]{\char126#1}

\newcommand{\lyxmathsym}[1]{\ifmmode\begingroup\def\b@ld{bold}
  \text{\ifx\math@version\b@ld\bfseries\fi#1}\endgroup\else#1\fi}

\usepackage{textcomp}
\usepackage{epstopdf}

\usepackage{amsfonts}

\usepackage{soul}

\pdfpageheight\paperheight
\pdfpagewidth\paperwidth

\@ifundefined{textcolor}{}{%
 \definecolor{BLACK}{gray}{0}
 \definecolor{WHITE}{gray}{1}
 \definecolor{RED}{rgb}{1,0,0}
 \definecolor{GREEN}{rgb}{0,1,0}
 \definecolor{BLUE}{rgb}{0,0,1}
 \definecolor{CYAN}{cmyk}{1,0,0,0}
 \definecolor{MAGENTA}{cmyk}{0,1,0,0}
 \definecolor{YELLOW}{cmyk}{0,0,1,0}
}

\usepackage{xcolor}\usepackage{soul}
\newcommand{\ket}[1]{\ensuremath{\left|#1\right\rangle}}

\definecolor{blue}{rgb}{0,0,1}
\definecolor{red}{rgb}{1,0,0}
\definecolor{green}{rgb}{0,1,0}

\makeatother

\begin{document}

\title{Hundred-hertz quantum circuit iteration rate in a reusable neutral-atom array}

\author{Liang~Chen}
\thanks{These authors contributed equally to this work.}
\affiliation{Laboratory of Quantum Information, University of Science and Technology of China, Hefei 230026, China.}
\affiliation{Anhui Province Key Laboratory of Quantum Network, University of Science and Technology of China, Hefei 230026, China}

\author{Wen-Yi~Zhu}
\thanks{These authors contributed equally to this work.}
\affiliation{Laboratory of Quantum Information, University of Science and Technology of China, Hefei 230026, China.}
\affiliation{Anhui Province Key Laboratory of Quantum Network, University of Science and Technology of China, Hefei 230026, China}

\author{Dong-Qi~Ma}
\thanks{These authors contributed equally to this work.}
\affiliation{Laboratory of Quantum Information, University of Science and Technology of China, Hefei 230026, China.}
\affiliation{Anhui Province Key Laboratory of Quantum Network, University of Science and Technology of China, Hefei 230026, China}

\author{Tian-Yang~Zhang}
\affiliation{Laboratory of Quantum Information, University of Science and Technology of China, Hefei 230026, China.}
\affiliation{Anhui Province Key Laboratory of Quantum Network, University of Science and Technology of China, Hefei 230026, China}

\author{Zi-Jie~Chen}
\affiliation{Laboratory of Quantum Information, University of Science and Technology of China, Hefei 230026, China.}
\affiliation{Anhui Province Key Laboratory of Quantum Network, University of Science and Technology of China, Hefei 230026, China}

\author{Yi-Chen~Zhang}
\affiliation{Laboratory of Quantum Information, University of Science and Technology of China, Hefei 230026, China.}
\affiliation{Anhui Province Key Laboratory of Quantum Network, University of Science and Technology of China, Hefei 230026, China}

\author{Hong-Jie~Fan}
\affiliation{Laboratory of Quantum Information, University of Science and Technology of China, Hefei 230026, China.}
\affiliation{Anhui Province Key Laboratory of Quantum Network, University of Science and Technology of China, Hefei 230026, China}

\author{Guang-Jie~Chen}
\affiliation{Laboratory of Quantum Information, University of Science and Technology of China, Hefei 230026, China.}
\affiliation{Anhui Province Key Laboratory of Quantum Network, University of Science and Technology of China, Hefei 230026, China}

\author{Qing-Xuan~Jie}
\affiliation{Laboratory of Quantum Information, University of Science and Technology of China, Hefei 230026, China.}
\affiliation{Anhui Province Key Laboratory of Quantum Network, University of Science and Technology of China, Hefei 230026, China}

\author{Wei-Zhou~Cai}
\affiliation{Laboratory of Quantum Information, University of Science and Technology of China, Hefei 230026, China.}
\affiliation{Anhui Province Key Laboratory of Quantum Network, University of Science and Technology of China, Hefei 230026, China}

\author{Tian-Cai~Zhang}
\affiliation{State Key Laboratory of Quantum Optics and Quantum Optics Devices, and Institute of Opto-Electronics, Shanxi University, Taiyuan 030006, China}
\affiliation{Collaborative Innovation Center of Extreme Optics, Shanxi University, Taiyuan 030006, China.}

\author{Luyan Sun}
\affiliation{Center for Quantum Information, Institute for Interdisciplinary Information Sciences, Tsinghua University, Beijing 100084, China}
\affiliation{Hefei National Laboratory, Hefei 230088, China.}

\author{Yan-Lei~Zhang}
\affiliation{Laboratory of Quantum Information, University of Science and Technology of China, Hefei 230026, China.}
\affiliation{Anhui Province Key Laboratory of Quantum Network, University of Science and Technology of China, Hefei 230026, China}
\affiliation{Hefei National Laboratory, Hefei 230088, China.}

\author{Xi-Feng~Ren}
\affiliation{Laboratory of Quantum Information, University of Science and Technology of China, Hefei 230026, China.}
\affiliation{Anhui Province Key Laboratory of Quantum Network, University of Science and Technology of China, Hefei 230026, China}

\author{Guang-Can~Guo}
\affiliation{Laboratory of Quantum Information, University of Science and Technology of China, Hefei 230026, China.}
\affiliation{Anhui Province Key Laboratory of Quantum Network, University of Science and Technology of China, Hefei 230026, China}

\author{Zhu-Bo~Wang}
\email{zbwang@ustc.edu.cn}
\affiliation{Laboratory of Quantum Information, University of Science and Technology of China, Hefei 230026, China.}
\affiliation{Anhui Province Key Laboratory of Quantum Network, University of Science and Technology of China, Hefei 230026, China}

\author{Ya-Dong~Hu}
\email{hyd1998@ustc.edu.cn}
\affiliation{Laboratory of Quantum Information, University of Science and Technology of China, Hefei 230026, China.}
\affiliation{Anhui Province Key Laboratory of Quantum Network, University of Science and Technology of China, Hefei 230026, China}

\author{Gang~Li}
\email{gangli@sxu.edu.cn}
\affiliation{State Key Laboratory of Quantum Optics and Quantum Optics Devices, and Institute of Opto-Electronics, Shanxi University, Taiyuan 030006, China}
\affiliation{Collaborative Innovation Center of Extreme Optics, Shanxi University, Taiyuan 030006, China.}

\author{Chang-Ling~Zou}
\email{clzou321@ustc.edu.cn}
\affiliation{Laboratory of Quantum Information, University of Science and Technology of China, Hefei 230026, China.}
\affiliation{Anhui Province Key Laboratory of Quantum Network, University of Science and Technology of China, Hefei 230026, China}
\affiliation{Hefei National Laboratory, Hefei 230088, China.}

\date{\today}

\begin{abstract}
\textbf{Neutral-atom quantum processors have rapidly advanced in scale and coherence, yet their practical performance remains constrained by limited quantum circuit iteration rates (qCIRs) and information throughput. Here we experimentally demonstrate a high-throughput neutral-atom system based on non-destructive readout and atom reuse. By integrating a chip-based photonic interface with a 10-qubit array, we implement non-destructive readout with a retention probability of 99.7\%, and further achieve a raw qCIR of 101\,Hz and a post-selected qCIR of 74.8\,Hz. More importantly, we verify a general throughput optimization methodology and obtain a normalized Fisher information rate of 57.7\,Hz, improving the achievable throughput by more than one order of magnitude compared with conventional methods. Our results establish a practical route toward high-throughput neutral-atom quantum processors.}
\end{abstract}
\maketitle

\noindent \textbf{\large{}Introduction}{\large\par}
\noindent The quest for practical quantum computing has spurred the development of various platforms capable of manipulating quantum bits (qubits) with high fidelity and scalability. Among these, neutral-atom arrays have emerged as promising candidates due to their intrinsic scalability in trapping and controlling large numbers of atoms with optical tweezers~\cite{Manetsch2025,Bluvstein2026,Evered2026,Radnaev2025,ma2023,Liu2023,Wangz2025,Ye2023,Zhang2026,Li2025,Zhu2026,Lin2026}. Recent advancements have led to the assembly of thousands of atoms in defect-free configurations~\cite{Manetsch2025,Holman2026,lin2024,Chiu2025,Wang2026} and two-qubit gate fidelities exceeding $99.9\%$~\cite{Evered2026}, both of which pave the way for exploring fault-tolerant quantum technologies. Despite this exciting progress, a critical challenge hinders the practical deployment of neutral-atom quantum processors. The qCIR, i.e., the number of complete circuits executed per second from initialization to readout, has remained orders of magnitude below that of a superconducting processor~\cite{Arute2019,Acharya2025}.


This limitation in qCIR arises from two aspects. First, a defect-free atom array takes a long time to prepare. Single atoms are loaded into optical tweezer arrays stochastically due to collisional blockade~\cite{Schlosser2001,Schlosser2002}, leading to a cost of a few hundred milliseconds for loading, imaging, and rearranging the array. For atom fluorescence detection with a camera, the exposure, frame-readout, and feedback control latencies take tens of milliseconds~\cite{Radnaev2025,Evered2026}. The second and more fundamental problem is the heating of an atom when its internal quantum state is read out. The detection is realized by the state-preserving scattering of photons, and every photon that carries information also imparts a recoil momentum kick. Collecting more photons improves the readout reliability but increases the escape probability, so the detection fidelity and atom loss are subject to a trade-off. Typically, fast but destructive detection on a cycling transition requires time-consuming re-preparation of the defect-free array; otherwise, detection durations exceeding tens of milliseconds with simultaneous cooling are needed to suppress atom loss. As a result, qCIRs in atom arrays have been limited to below ten hertz over the past decade, even when atoms are reused~\cite{Radnaev2025,Bluvstein2026}.

\begin{figure*}
\begin{centering}
\includegraphics[width=0.9\linewidth]{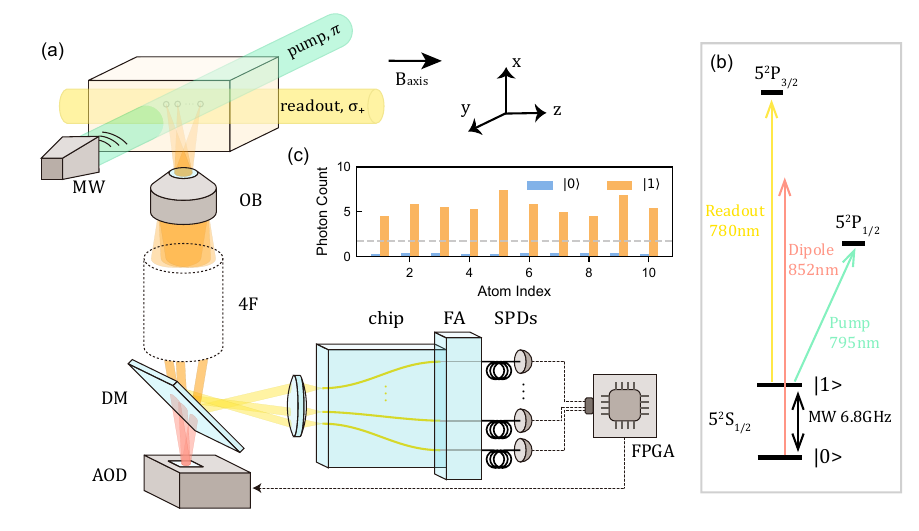}
\par\end{centering}
\caption{Photonic-chip-enabled parallel site-resolved readout architecture for a neutral-atom array. 
(a) Experimental configuration. Ten $^{87}\mathrm{Rb}$ atoms are trapped in a one-dimensional tweezer array generated by an acousto-optic deflector (AOD). Global $\pi$-polarized microwave (MW), $\pi$-polarized optical-pumping, and $\sigma^{+}$-polarized readout fields are used for qubit state initialization, gate operation, and readout. Atomic fluorescence collected by the objective (OB) is routed through a laser-written three-dimensional (3D) optical mapping chip, whose output pitch is matched to a fiber array (FA), and detected by independent single-photon detectors (SPDs). Detection events are processed in real time by a field-programmable gate array (FPGA). DM, dichroic mirror. 
(b) Relevant $^{87}\mathrm{Rb}$ energy levels and optical and microwave couplings. The qubit states $\ket{1}$ and $\ket{0}$ are encoded in the $5S_{1/2}$ hyperfine manifold. 
(c) Representative site-resolved photon counts after initialization in $\ket{1}$ and $\ket{0}$, acquired with a readout duration $\tau_\mathrm{ro}=$1.5\,ms. The dashed line denotes the photon-count discrimination threshold.}

\label{Fig1}
\end{figure*}

Here, we demonstrate a high-throughput neutral-atom quantum processor operating at a hundred hertz. By interfacing the atom array with an integrated 3D photonic chip~\cite{Ma2026}, we establish a direct qubit-to-detector correspondence that enables state-resolved, non-destructive, and fast readout~\cite{kwon2017,martinez-dorantes2017,martinez-dorantes2018,shea2020,nikolov2023,chow2023,WJ2025}, and thereby significantly enhance the qCIR by mitigating the need for reloading and rearrangement between circuit executions. Consequently, we achieve a raw qCIR exceeding 100\,Hz and a post-selected qCIR of 74.8\,Hz. The resulting normalized Fisher information rate is 57.7\,Hz, showing an improvement of more than an order of magnitude over a destructive-readout approach even if its readout were perfect. A single-qubit-gate randomized benchmarking experiment, which gives an average gate fidelity of 99.93\%, is performed in 13.2 minutes for each data point instead of several hours~\cite{nikolov2023}. These results present the throughput-oriented optimization~\cite{Chen2025} of atomic quantum processors in experiment, and open a route toward kHz-qCIR operation that will further improve the throughput of this platform. 


\noindent \textbf{\large{}Results}{\large\par}

\noindent
Figure~\ref{Fig1}(a) shows a schematic of our experimental setup for the $^{87}$Rb atomic array, with the relevant energy levels depicted in Fig.~\ref{Fig1}(b). The quantization axis is defined by a $4\,\mathrm{G}$ magnetic field $B_\mathrm{axis}$ oriented along the axial direction of the vacuum cell, i.e., the $z$ direction. Ten dipole traps, formed by $\pi$-polarized laser beams at $852\,\mathrm{nm}$ and arranged in a one-dimensional array along the $z$ direction, serve to load single atoms that serve as qubits. The $\ket{1}$ state of the qubit is encoded in $5S_{1/2},F=2,m_F=0$ state, while $\ket{0}$ state is encoded in $5S_{1/2},F=1,m_F=0$ state. An optical pumping beam, propagating along the $y$ direction, drives the $5S_{1/2}, F=2 \rightarrow 5P_{1/2}, F'=2$ transition with $\pi$-polarized light for qubit state initialization of the array. A microwave horn is mounted along the $-y$ direction to drive the transition between two qubit states. Qubit state readout is performed using a laser beam propagating along the $z$ direction and resonant with the $5S_{1/2}, F=2, m_F=2 \rightarrow 5P_{3/2}, F'=3, m_F'=3$ cycling transition. All of the aforementioned control fields are global, addressing the entire atomic array, and are controlled by a single FPGA~\cite{Hu2026}.

\begin{figure*}
\begin{centering}
\includegraphics[width=0.9\linewidth]{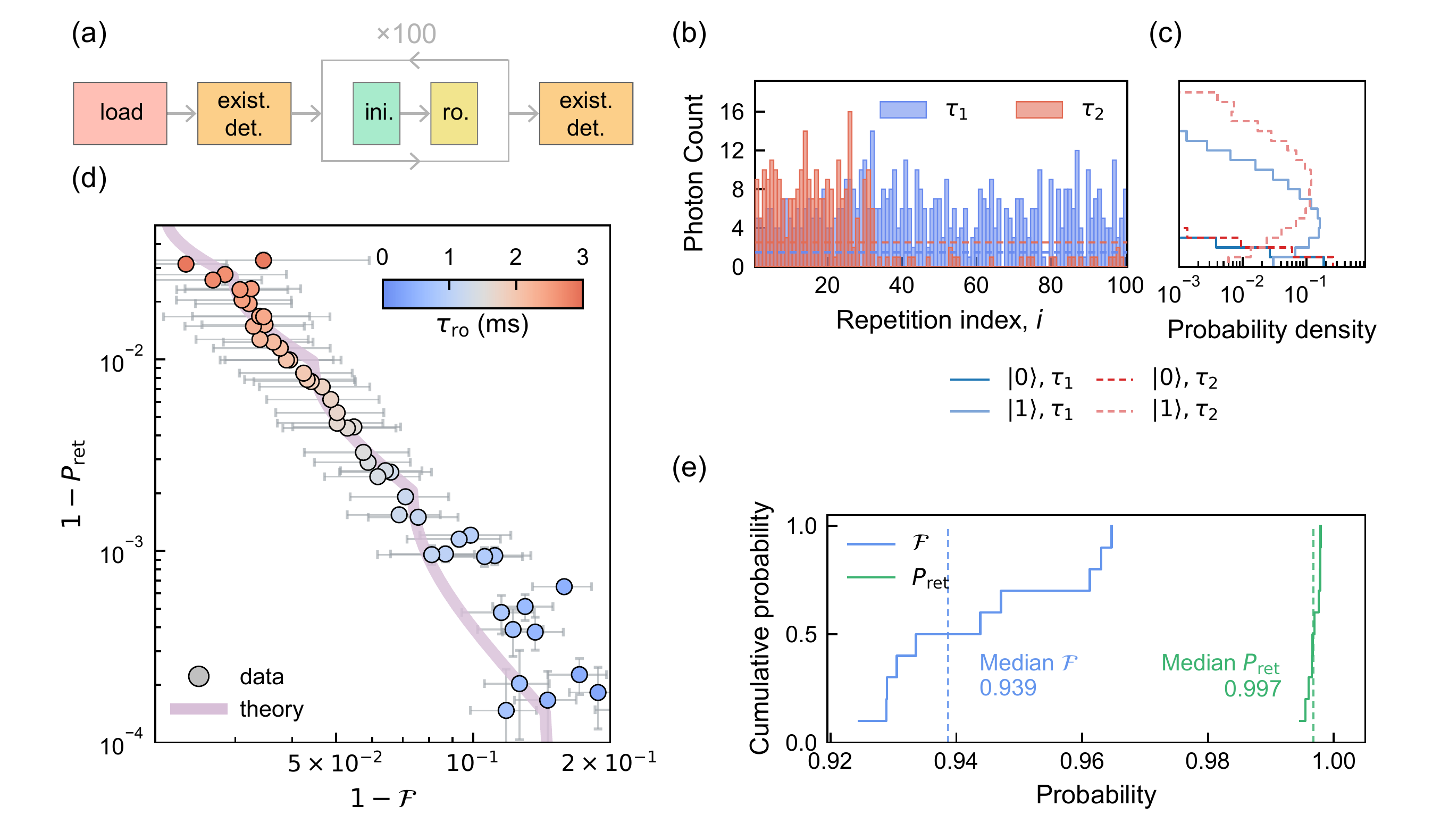}
\par\end{centering}
\caption{Non-destructive readout and atom reuse.
(a) Benchmark sequence of non-destructive readout. After atom loading and an initial fluorescence-based existence detection (exist. det.), qubit initialization (ini.) including cooling and qubit state initialization and readout (ro.) are repeated 100 times, followed by a final existence detection. The loading time, each existence-detection time, and the duration of one repeated cycle are $200\,\mathrm{ms}$, $45\,\mathrm{ms}$, and $7\,\mathrm{ms}$, respectively.
(b) Representative readout signal datasets over 100 repetitions for two readout durations, $\tau_1=1.5\,\mathrm{ms}$ and $\tau_2=2.5\,\mathrm{ms}$.
(c) Corresponding readout signal distributions for $\ket{1}$ and $\ket{0}$ states.
(d) Trade-off between the array-averaged readout infidelity $1-\mathcal F$ and the single-shot atom-loss probability $1-P_\mathrm{ret}$ as the readout duration $\tau_\mathrm{ro}$ is varied. The color scale denotes $\tau_\mathrm{ro}$; horizontal and vertical error bars represent one standard error. The purple curve is the model prediction obtained using independently characterized experimental parameters.
(e) Cumulative distributions of the site-resolved readout fidelity $\mathcal F$ and single-shot retention probability $P_\mathrm{ret}$ at $\tau_\mathrm{ro}=1.5\,\mathrm{ms}$. Dashed vertical lines mark the median values, $\mathcal F=93.9\%$ and $P_\mathrm{ret}=99.7\%$.}
\label{Fig2}
\end{figure*}

The core of this ``volcano" architecture~\cite{Ma2026} is the optical channel mapping (OCM) chip fabricated by laser direct writing. This chip contains ten single-mode waveguides corresponding to the atom array. The waveguides are arranged in a dense parallel array on one end (left side in Fig.~\ref{Fig1}(a)) and a sparse parallel array on the other (right side in Fig.~\ref{Fig1}(a)). The waveguide pitch at the densely packed end is $50\,\mu\mathrm{m}$, which matches the $4.5\,\mu\mathrm{m}$ atom array spacing after optical conversion. At this end, the waveguide modes are collimated by a lens and are aligned one-to-one with the dipole-trap beams generated by the AOD at a DM. The coupled modes are subsequently expanded through a series of lenses and focused by an objective to form a one-dimensional array inside the vacuum cell. The sparsely packed end of the chip, with a $127\,\mu\mathrm{m}$ pitch, is directed towards the fluorescence collection apparatus. It is first butt-coupled to a standardized fiber array with a matching pitch, and the fluorescence is subsequently guided through these fibers to a SPD array. Finally, the SPD array converts the atomic fluorescence into electronic signals, which are transmitted to an FPGA for further processing. This volcano OCM chip introduces approximately 3\,dB of insertion loss and crosstalk on the order of 0.1\%, as characterized following Ref.~\cite{Ma2026}.

We first implement an atom-reuse strategy using the sequence shown in Fig.~\ref{Fig2}(a) and perform non-destructive-readout benchmarking under repeated operations. After the atom-loading stage, an atom existence detection is performed, followed by 100 consecutive benchmark cycles, and concluding with another existence detection. The atom existence detection is realized by applying the MOT lasers to the array to check whether each atom still exists in its trap or has been lost. During the readout stage, we increase the trap depth to $5\,\mathrm{mK}$, and alternate between the trap and the readout laser at a frequency of 1\,MHz and a duty cycle of 0.5 to suppress atom loss. The loading stage costs $200\,\mathrm{ms}$, each existence detection costs $45\,\mathrm{ms}$, and a single benchmark cycle costs $7\,\mathrm{ms}$. Consequently, under this sequence our system achieves a raw qCIR $\mathcal{R}_\mathrm{raw}=101\,\mathrm{Hz}$. This benchmark primarily concerns the intrinsic trade-off between readout fidelity $\mathcal{F}$ and retention probability $P_\mathrm{ret}$. Scattering more photons during a single readout stage yields higher single-shot readout fidelity $\mathcal{F}$, but simultaneously result in a higher final atom temperature and therefore a lower retention probability $P_\mathrm{ret}$. Figure~\ref{Fig2}(b) shows two typical datasets of readout signal for 100 cycles. The red bars correspond to a longer readout duration $\tau_2=2.5\,\mathrm{ms}$, for which atoms in $\ket{1}$ exhibit higher fluorescence counts and thus higher $\mathcal{F}$, yet are more prone to loss during the cycles. The blue bars correspond to a shorter $\tau_1=1.5\,\mathrm{ms}$, yielding lower $\mathcal{F}$ but allowing the atoms to survive through more cycles. The corresponding signal histograms are presented in Fig.~\ref{Fig2}(c).

\begin{figure*}
\begin{centering}
\includegraphics[width=0.8\linewidth]{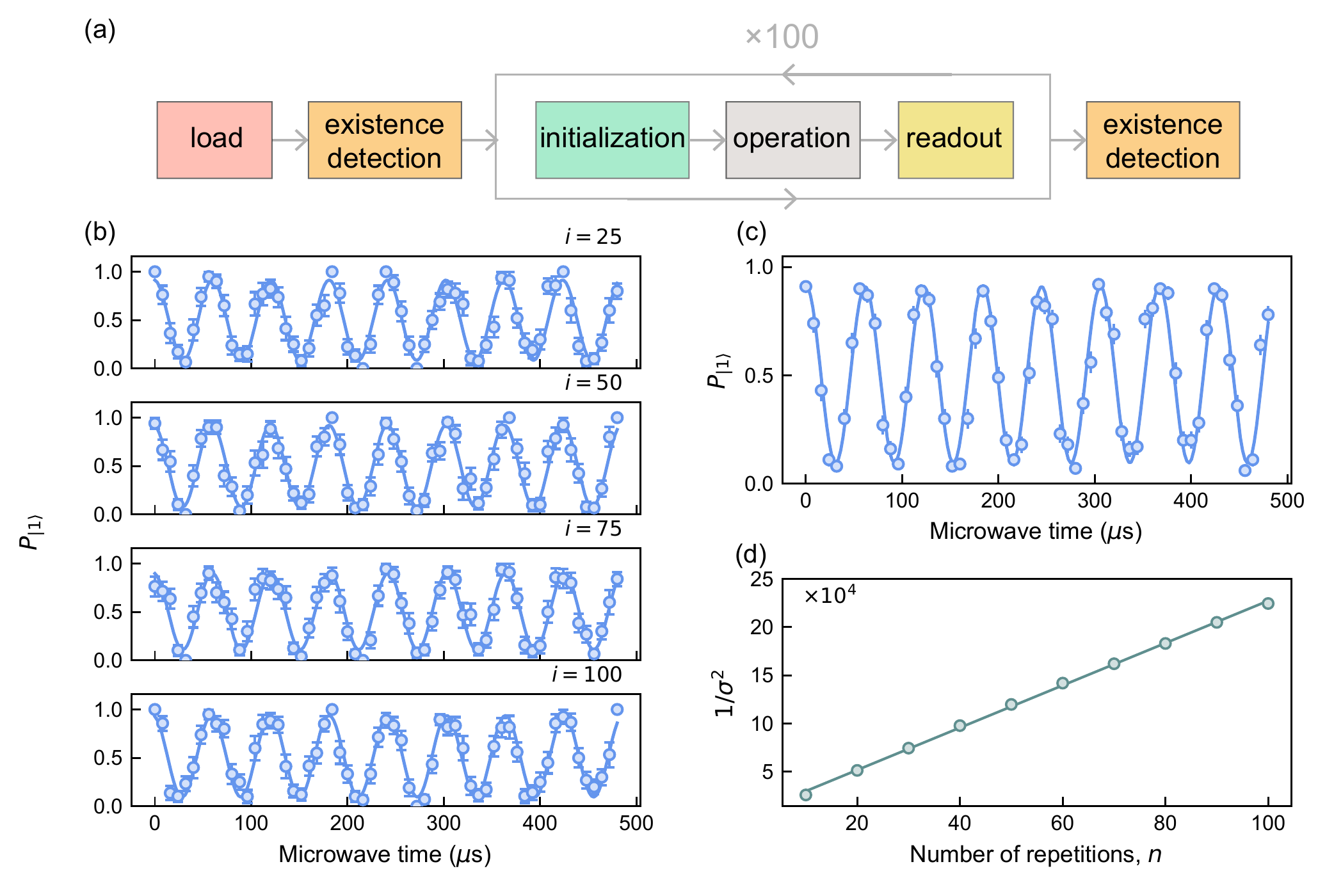}
\par\end{centering}
\caption{Stable information accumulation during repeated circuit execution.
(a) Sequence used for Rabi-oscillation sampling. A microwave-driven single-qubit gate operation (op.) is inserted between qubit initialization and readout in each of the 100 reuse cycles.
(b) Rabi oscillations reconstructed separately from the 25th, 50th, 75th, and 100th repetitions. Points show the measured $\ket{1}$ state population $P_{\ket{1}}$, and solid curves are independent fits to a sinusoid function. Error bars represent one standard error.
(c) Rabi oscillation obtained by combining data from all 100 repetitions. Points show the measured $\ket{1}$ state population $P_{\ket{1}}$, and solid curve is a fit to a sinusoid function. Each data point corresponds to one complete sample sequence and 0.99\,s.
(d) Inverse variance $1/\sigma^{2}$ of the population as a function of the number $n$ of included repetitions. The solid line is a linear fit. The observed linear scaling indicates statistically consistent information accumulation over the repeated cycles.}
\label{Fig3}
\end{figure*}

By finely scanning $\tau_\mathrm{ro}$, we map out the trade-off between the array-averaged readout infidelity $1-\mathcal{F}$ and the array-averaged single-shot atom loss probability $P_\mathrm{loss}=1-P_\mathrm{ret}$, as plotted in Fig.~\ref{Fig2}(d). The purple solid line is the theoretical curve calculated according to the model introduced in Ref.~\cite{Chen2025}, starting from a series of independently calibrated experimental parameters (such as initial atom temperature, trap depth, scattering rate, and collection efficiency). For an optimized state detection duration of $1.5\,\mathrm{ms}$, the distributions of $\mathcal{F}$ and $P_\mathrm{ret}$ for 10 atoms are plotted as cumulative distribution function (CDF) curves in Fig.~\ref{Fig2}(e). The median readout fidelity and retention probability are found to be 93.9\% and 99.7\%, respectively. The selection of this set of parameters sacrifices fidelity to a certain extent, but yields a higher qCIR and an overall superior throughput. 
Constrained by the current hardware, we adopt a simple but conservative post-selection: the data from these 100 circuit samples are retained only if both the initial and final detections confirm that the atom is present. The resulting effective qCIR after post-selection is $\mathcal{R}_\mathrm{pos}=\mathcal{R}_\mathrm{raw}\times P_\mathrm{ret}^{100}=74.8\,\mathrm{Hz}$.


Building on the combination of nondestructive readout and atom reuse strategy, we substantially increase the system’s sampling rate by incorporating the execution of specific circuits into each cycle, including the state initialization, qubit operations, and state readout, as shown in Fig.~\ref{Fig3}(a). To verify the consistency of the repeated sampling, we measure microwave-driven Rabi oscillations between the $\ket{0}$ and $\ket{1}$ states and analyze their uniformity. Figure~\ref{Fig3}(b) shows several Rabi oscillation curves obtained by processing data selected from only the 25th, 50th, 75th, and 100th repetitions of the 100-cycle sequence, along with their respective fitted curves. The whole dataset is collected from one specific atom channel, and each data point corresponds to around 25 samples. It can be seen that the primary characteristics of these oscillations, including their upper and lower bounds, frequency, and phase, are similar. Figure~\ref{Fig3}(c) presents the result obtained by sampling only one complete sequence (0.99\,s) from the same atom channel and superimposing the data from the 100 cycles within that sequence. As expected, due to the large sample size of 100 and the high data consistency, the error bars on the data points in Fig.~\ref{Fig3}(c) are reduced by an order of magnitude. To quantitatively analyze the consistency of the 100 repetitions, we perform the following analysis: we pick the data of the first $n$ repetitions on the same atom channel, examine the resulting Rabi fit parameters, and derive the inverse of the population variance (1/$\sigma^2$) as a function of $n$. The results are plotted in Fig.~\ref{Fig3}(d). As the number of included repetitions $n$ increases, the inverse variance exhibits a linear growth. This demonstrates that the data from the different repetitions are highly consistent, contribute the same Fisher information, and can be combined without introducing additional bias.

\begin{figure*}
\begin{centering}
\includegraphics[width=0.8\linewidth]{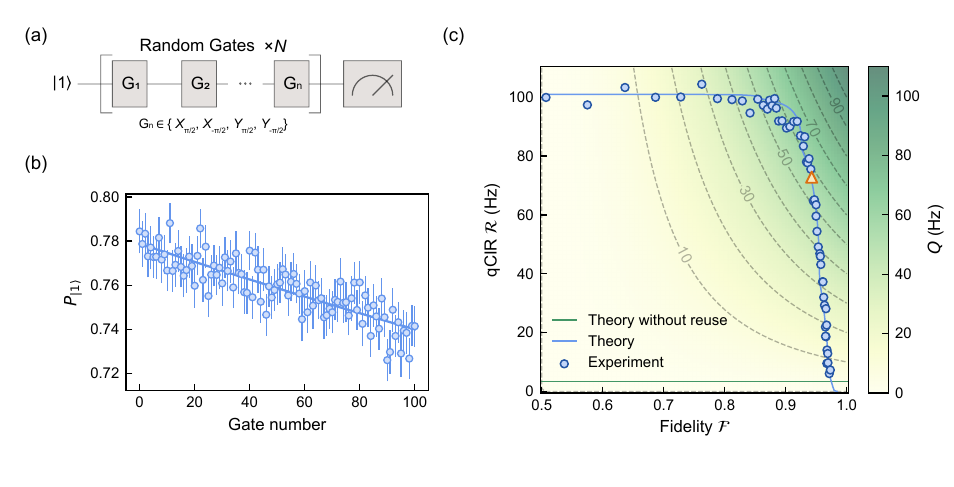}
\par\end{centering}
\caption{Rapid single-qubit randomized benchmarking and optimization of the normalized Fisher information rate.
(a) Single-qubit randomized-benchmarking circuit. The atom is prepared in $\ket{1}$ and then subjected to $n$ randomly sampled gates $G_n\in\{X_{\pi/2},X_{-\pi/2},Y_{\pi/2},Y_{-\pi/2}\}$ including 1 or 2 recovery gates chosen such that the ideal final state is $\ket{1}$, before state readout.
(b) Measured $\ket{1}$ state probability $P_{\ket{1}}$ as a function of gate number $n$. Each data point is sampled over 593 random sequences, and averaged over 10 atoms. Error bars represent one standard error, and the solid curve is an exponential fit, yielding an average fidelity of 99.93\%.
(c) Measured quantum circuit iteration rate after post-selection $\mathcal{R}_\mathrm{pos}$ as a function of readout fidelity $\mathcal{F}$ as $\tau_\mathrm{ro}$ is varied. The blue solid curve is the model prediction obtained using independently characterized experimental parameters, while the green solid curve is derived from the same parameter set but without applying the atom-reuse strategy. The background color and dashed contours show the effective normalized Fisher information rate
$Q=\mathcal{R}_\mathrm{pos}(2\mathcal{F}-1)^2$. The triangle marks the operating point used for the benchmarking measurement, $(\mathcal{F},\mathcal{R}_\mathrm{pos},Q)=(93.9\%,74.8\,\mathrm{Hz},57.7\,\mathrm{Hz})$.}
\label{Fig4}
\end{figure*}


Employing the high-throughput atomic quantum processor, we implement single-qubit randomized benchmarking (RB) on the 10-qubit array. The quantum circuit for RB is depicted in Fig.~\ref{Fig4}(a), which uses a gate set of \{$X_{\pi/2}, X_{-\pi/2}, Y_{\pi/2}, Y_{-\pi/2}$\}. The array-averaged characterization results are shown in Fig.~\ref{Fig4}(b), in which each data point is sampled over 593 random sequences, and averaged over 10 atoms. This sample size corresponds to an acquisition time of 13.2 minutes for each data point. A fit yields an average single-qubit gate fidelity of 99.93\%. 

Furthermore, we verify the method for optimizing system throughput, which has been introduced in our previous theoretical work~\cite{Chen2025}. This methodology focuses on the trade-off between $\mathcal{F}$ and $\mathcal{R}_\mathrm{pos}$ associated with the total number of scattered photons during a single readout stage. We map this trade-off by further processing the data corresponding to Fig.~\ref{Fig2}(d). The experimental data are plotted as points in Fig.~\ref{Fig4}(c). The blue solid line is the theoretical curve calculated according to the model introduced in Ref.~\cite{Chen2025}, starting from a series of independently calibrated experimental parameters, while the green solid line is the theoretical curve derived from the same parameter set but assuming that the atom-reuse strategy is not used.

The normalized Fisher information rate $Q$ is introduced as a comprehensive figure of merit, because it reflects the amount of information the system can obtain per unit time, given by the expression $Q = \mathcal{R}_\mathrm{pos}(2\mathcal{F}-1)^2$~\cite{wang2022,meyer2021,liu2020,Chen2025}. In Fig.~\ref{Fig4}(c), the value of $Q$ corresponding to each ($\mathcal{F}$, $\mathcal{R}_\mathrm{pos}$) point is plotted as a color map in the background, with contours also shown as gray dashed lines. The readout parameters adopted in this work yield a $\mathcal{F}$ of 93.9\% and a $\mathcal{R}_\mathrm{pos}$ of 74.8\,Hz, corresponding to a $Q$ of 57.7\,Hz. For the same system operated without non-destructive readout and atom reuse strategy, the system’s $Q$ would be severely limited by the qCIR. Even with an ideal readout fidelity $\mathcal{F}=1$, $Q$ could only reach 3.37\,Hz. This result represents an improvement of more than one order of magnitude in $Q$ compared with conventional systems. Moreover, it proves that we demonstrates that the methodology is broadly applicable, which can enable more than an order of magnitude improvement in $Q$ without replacing the system hardware, simply by adjusting the experimental parameters and sequence strategy.

\noindent \textbf{\large{}Discussion}{\large\par}
\noindent 
In conclusion, we have demonstrated that a neutral-atom quantum processor can execute quantum circuits at an iteration rate of 100~Hz. The achieved qCIR is determined by the 7\,ms duration of each circuit iteration cycle, which consists of a cooling process of about 5\,ms and a readout process of 1.5\,ms. We anticipate that both the cooling and readout durations can be reduced to $\sim 0.1\,\mathrm{ms}$ by improving the photon collection efficiency~\cite{Chen2024,Crain2019,Su2025,MuziFalconi2025} and by adopting advanced readout schemes that suppress recoil heating~\cite{chow2023,Su2025,MuziFalconi2025,Yokoyama2026}, offering a route toward neutral-atom quantum processors operating at kilohertz qCIR. Leveraging the capability of the ``volcano" architecture for arbitrary site-selective addressing~\cite{Ma2026}, we will extend the high-throughput quantum circuits to two-dimensional qubit arrays and implement dynamical quantum circuits with mid-circuit measurements and real-time feedback~\cite{deist2022,Graham2023,Norcia2023,Huie2023}.

\smallskip{}
\begin{acknowledgments}
We acknowledge Xu-Liang Zhang from Hangzhou Biaozhang Electronics for assistance. This work was funded by the National Key R\&D Program (Grant Nos. 2021YFA1402004 and 2021YFA1402002), the National Natural Science Foundation of China (Grant Nos.~92465201, T2325022 and U23A2074), and the Quantum Science and Technology-National Science and Technology Major Project (Grant Nos. 2021ZD0303200 and 2021ZD0301500). This work was also supported by  the Natural Science Foundation of Anhui
Province (Grant No.~2408085QA017), the Fundamental Research Funds for the Central Universities, and the USTC Research Funds of the Double First-Class Initiative. This work was partially carried out at the Supercomputing Center of USTC and the USTC Center for Micro and Nanoscale Research and Fabrication.
\end{acknowledgments}

\end{document}